\documentclass{svproc}
\usepackage{url}

\usepackage{graphicx}

\begin{document}
\mainmatter              % start of a contribution
\title{Observable Primordial Gravitational Waves from Cosmic Inflation}
\titlerunning{Inflation in Metric-Affine Quadratic Gravity}
% abbreviated title (for running head)
\author{Konstantinos Dimopoulos\inst{1}}
\authorrunning{Konstantinos Dimopoulos} % abbreviated author list (for running head)
\tocauthor{Konstantinos Dimopoulos}
\institute{Consortium for Fundamental Physics,\\ Physics Department, Lancaster University, Lancaster LA1 4YB, UK\\
\email{k.dimopoulos1@lancaster.ac.uk},\\ WWW home page:
\texttt{https://www.lancaster.ac.uk/physics/about-us/people/konstantinos-dimopoulos}}

\maketitle              % typeset the title of the contribution

\begin{abstract}
I will review briefly how inflation is expected to generate a stochastic background of primordial
gravitational waves (GWs). Then, I will discuss how such GWs can be enhanced by a stiff period
following inflation, enough to be observable. I will present examples of this in the contact of hybrid
inflation with $\alpha$-attractors, or a period of hyper-kination in Palatini gravity.
\end{abstract}
\section{Introduction}
%	Cosmic Inflation}
%
The history of the Universe requires special initial conditions which are arranged by cosmic inflation. In a nutshell, cosmic inflation can be defined as a period of accelerated expansion in the Early Universe. Inflation results in the Universe being large, spatially flat and uniform, in accordance to observations
\cite{guth}. Inflation also generates the primordial density perturbations (PDPs), which are necessary for galaxies and galaxy clusters to form \cite{staro}. 
The PDPs reflect themselves on the cosmic microwave  background (CMB) radiation, through the Sachs-Wolfe effect. Agreement with CMB observations is spectacular, when inflation is quasi-de Sitter and space expands exponentially \cite{planck}. 

However, there is another generic prediction of inflation beyond the acoustic peaks observed in the CMB, namely the generation of a stochastic spectrum of primordial gravitational waves (GWs)
\cite{GWinf}. This prediction will soon be tested, either indirectly, through CMB polarisation observations, or directly through interferometers (Advanced LIGO, LISA). 

How does inflation produce these GWs? Below, I attempt a brief overview. I use natural units with $c=\hbar=1$ and $8\pi G=m_P^{-2}$, with $m_P=2.43\times 10^{18}\,$GeV being the reduced Planck mass. The signature of the metric is positive.

\section{Particle production of gravitational waves during cosmic inflation}
Following the recipe of linearised gravity, we consider a comoving perturbation of the metric $h_{ij}$, such that the line element of spatially flat FRW spacetime is
\begin{equation}
	{\rm d}s^2=a^2(\tau)[-{\rm d}\tau^2+(\delta_{ij}+h_{ij})
	{\rm d}x^i{\rm d}x^j]\,,
\end{equation}
where $a(\tau)$ is the scale factor of the Universe as a function of conformal time $\tau$ and $\delta_{ij}$ is equal to one when $i=j$ and to zero otherwise, with \mbox{$i,j=1,2,3$}. The metric perturbation is symmetric \mbox{$h_{ij}=h_{ji}$}, traceless
\mbox{$h_i^i=0$} and transverse \mbox{$\nabla_i h^{ij}=0$}. This means that it is corresponds to two degrees of freedom \mbox{(6-1-3=2)}, which are the two polarisations, $\oplus$ and $\otimes$, of the GWs.

Thus, by Fourier transform, we can write
\begin{equation}
	h_{ij}(\tau, {\bf x})=\sqrt{16\pi G}
	\int\frac{{\rm d}^3k}{(2\pi)^{3/2}}
	h_{ij}(\tau, {\bf k})\,e^{i{\bf k\cdot x}}\,,
\end{equation}
with 
\begin{equation}
	h_{ij}(\tau, {\bf k})=\sum_{s=\oplus,\otimes} h_k^s
	e_{ij}^s({\bf k})\,,
\end{equation}
where $e_{ij}^2$ is symmetric \mbox{$e_{ij}^2=e_{ji}^s$}, traceless \mbox{$e_i^{s\;i}=0$} and transverse
\mbox{$k^ie_{ij}^s=0$}, with $k^i$ being the 3-wavevector of the GWs.

The second-order GW action is 
\begin{equation}
	S_{\rm GW}=\frac{1}{64\pi G}
	\int{\rm d}^4x\sqrt{-g}\, g^{\mu\nu}
	\partial_\mu h_{ij}\,
	\partial_\nu h^{ij}\,,
	\label{SGW}
\end{equation}
where $g$ is the determinant of the metric $g_{\mu\nu}$ and \mbox{$\mu,\nu=0,1,2,3$}.
From the above action one obtains the equation of motion (EoM), which reads
\begin{equation}
	h_{ij}''+2\frac{a'}{a}h_{ij}'-\nabla^2
	h_{ij}=0\;\Rightarrow\;
	{h_k^s}''+2\frac{a'}{a}{h_k^s}'+k^2 h_k^s=0
	\,,
	\label{EoM}
\end{equation}
where the prime denotes derivative with respect to conformal time.

The above equations show that the metric perturbation polarisations behave as free massless scalar fields $\psi_k^s(\tau)$ which can be written as \mbox{$h_k^s(\tau)=\sqrt{16\pi G}\psi_k^s(\tau)$}. To study particle production, we introduce the Muchanov-Sasaki variable
%\mbox
{$v_k^s(\tau)\equiv a(\tau)\psi_k^s(\tau)$}. In terms of this variable, the EoM becomes the well-known
Muchanov-Sasaki equation %\cite{MukhanovSasaki}
\begin{equation}
	{v_k^s}''+\left(k^2-\frac{a''}{a}\right)v_k^s=0\,,
	\label{MukhSas}
\end{equation}
where \mbox{$v_k^2=ah_k^s m_P/\sqrt 2$}.

To proceed we quantize the metric perturbations (second quantization) be expanding them in terms of creation and annihilation operators as
\begin{equation}
	v^s(\tau,{\bf x})=
	\int\frac{{\rm d}^3k}{(2\pi)^{3/2}}
	\left[v_k^s\hat a_{\bf k}^s\,
	e^{i{\bf k\cdot x}}+
	(v_k^s)^*\hat a_{\bf k}^{s\;\dagger}\,
	e^{-i{\bf k\cdot x}}
	\right]\,,
\end{equation}
where $\hat a_{\bf k}^s$ and $\hat a_{\bf k}^{s\;\dagger}$ are creation and annihilation operators respectively, which satisfy the algebra
\begin{equation}
	[\hat a_{\bf k}^s,\hat a_{\bf q}^{r\;\dagger}]=\delta^{sr}\delta^{(3)}({\bf q}-{\bf k})\quad{\rm and}\quad
	[\hat a_{\bf k}^s,\hat a_{\bf q}^r]=
	[\hat a_{\bf k}^{s\;\dagger},\hat a_{\bf q}^{r\;\dagger}]=0\,,
\end{equation}
where the value of $\delta^{sr}$ 
%$\{\delta^{(3)}({\bf q}-{\bf k})\}$ 
is unity when $s=r$ 
%$\{{\bf q}=\bf k\}$ 
or zero otherwise.

Inserting the above in Eq.~(\ref{MukhSas}), the solution for the mode functions is
\begin{equation}
	v_k^s(\tau)=\frac{1}{\sqrt{2k}}
	\left(1-\frac{i}{k\tau}\right)
	e^{-ik\tau}\,.
\end{equation}
The above solution, in the subhorizon limit
$-k\tau\rightarrow+\infty$ becomes \mbox{$v_k^s\rightarrow e^{-ik\tau}/\sqrt{2k}$}, which is the well-known Bunch-Davis vacuum \cite{BunchDavis}. 
In the superhorizon limit {$-k\tau=\frac{k}{aH}\rightarrow 0$}, the above solution becomes
\mbox{$v_k^s\rightarrow \frac{i}{\sqrt{2k}}\frac{aH}{k}$}, where 
$H$ is the Hubble parameter. Thus, in the superhorizon limit we find
\begin{equation}
	h_k^s=\frac{\sqrt 2\,v_k^s}{a m_P}=
	\frac{iH}{m_P k^{3/2}}\;\Rightarrow\;
	|h_k^s|^2=\frac{H^2}{m_P^2k^3}\approx
	{\rm constant}\,,
\end{equation}
where we considered that $H\approx\,$constant
in quasi-de Sitter inflation.

Therefore, we see that on superhorizon scales inflation generates a spectrum of primordial GWs. The value of their spectrum is
\begin{equation}
	{\cal P}_h(k)=\frac{k^3}{2\pi^2}
	\langle h_{ij}(k)\,h^{ij}(k)\rangle=
	\frac{k^3}{\pi^2}\sum_{s=\oplus,\otimes}
	|h_k^s|^2=\frac{2H^2}{\pi^2m_P^2}=64\pi G\left(\frac{H}{2\pi}\right)^2\,.
	\label{Ph}
\end{equation}
The scalar perturbations corresponding to the PDPs have spectrum
\begin{equation}
	{\cal P}_\zeta(k)=\frac{H^2}{8\pi^2\epsilon m_P^2}\,,
	\label{Pz}
\end{equation}
where \mbox{$\epsilon\equiv-\dot H/H^2$}, with the dot denoting derivative with respect to the cosmic time. 
%Dividing Eq.~(\ref{Ph}) with Eq.~(\ref{Pz}) 
From the above,
we find the consistency equation \mbox{$r\equiv{\cal P}_h/{\cal P}_\zeta=16\epsilon$}, which can be tested in the near future. At the moment, the CMB observations impose only an upper bound on $r$: \mbox{$0\leq r<0.036$} \cite{Keck}.

%Despite their quantum origin by virtue of the Bunch-Davis vacuum, the metric perturbations at superhorizon scales are considered ``classicalised'', through a process called quantum decoherence. The rationale is that, while a state in the quantum vacuum has \mbox{$[\psi,\dot\psi]=i\hbar$} and is originally a blob in the origin of phase space, when taken over by the inflationary expansion of space as it is pushed out of the horizon, it becomes "squeezed" such that \mbox{$\psi,\dot\psi\gg[\psi,\dot\psi]$}. The commutator is therefore negligible and the state can be considered approximately classical with \mbox{$[\psi,\dot\psi]\rightarrow 0$}.

\section{Density of the gravitational waves}
In view of Eq.~(\ref{SGW}), the energy-momentum tensor of the GWs is
\begin{equation}
	T_{\mu\nu}^{\rm GW}=-\frac{2}{\sqrt{-g}}
	\frac{\delta S_{\rm GW}}{\delta g^{\mu\nu}}=
	\frac{\langle\nabla_\mu h_{ij}\nabla_\nu h^{ij}\rangle}{32\pi G}\,,
\end{equation}
from which we can read the density \mbox{$\rho_{\rm GW}=a^{-2}T_{00}$}.
%\begin{equation}
%	\rho_{\rm GW}=-T_0^0=a^{-2}T_{00}=
%	\frac{1}{32\pi G}\frac{|h_{ij}'|^2+|\nabla h_{ij}|^2}{a^2}\,,
%\end{equation}
%where we considered Eq.~(\ref{EoM}). 
Using  Eq.~(\ref{EoM}),
%the equation of motion 
we can write
\begin{equation}
	\langle\rho_{\rm GW}\rangle=\int_0^{+\infty}
	\frac{k^3}{2\pi^2}
	\sum_{s=\oplus,\otimes}
	\frac{|{h_k^s}'|^2+k^2|h_k^s|^2}{a^2}
	\,{\rm d}(\ln k)\,.
	\label{<rho>}
\end{equation}	
Similarly, we find the isotropic pressure \mbox{$p_{\rm GW}=a^{-2}T_i^i$}, for which
\begin{equation}
	\langle p_{\rm GW}\rangle=\int_0^{+\infty}
	\frac{k^3}{2\pi^2}
	\sum_{s=\oplus,\otimes}
	\frac{|{h_k^s}'|^2-\frac13 k^2|h_k^s|^2}{a^2}
	\,{\rm d}(\ln k)\,.
	\label{<p>}
\end{equation}	
The above integrals are dominated by the subhorizon limit \mbox{$k\gg aH$}. In this limit we have \mbox{$h_k^s\propto e^{-ik\tau}$}, which implies 
\mbox{${h_k^s}'=-ik\,h_k^s\Rightarrow 
	|{h_k^s}'|^2=k^2|h_k^s|^2$}. 
Therefore, for the barotropic parameter we find
\begin{equation}
	w_{_{\rm GW}}=\frac{p_{\rm GW}}{\rho_{\rm GW}}=\frac{\frac23 k^2|h_k^s|^2}{2k^2|h_k^s|^2}=\frac13\,.
	\label{wGW}
\end{equation}
Thus, we find that the density of the gravitational waves redshifts as radiation with the Universe expansion \mbox{$\rho_{\rm GW}\propto a^{-3(1+w_{_{\rm GW}})}=a^{-4}$}.

In view of the above, the density parameter of the GWs per logarithmic momentum interval is
\begin{equation}
	\Omega_{\rm GW}(k)\equiv\frac{1}{\rho_c}
	\frac{{\rm d}\rho_{\rm GW}}{{\rm d}\ln k}=\frac{k^3}{2\pi^2}\frac{8\pi G}{3H^2}
	\sum_{s=\oplus,\otimes}
	\frac{|{h_k^s}'|^2+k^2|h_k^s|^2}{a^2}
	\,.
	\label{OGW}
\end{equation}
where \mbox{$\rho_c=3H^2/8\pi G$} is the critical density. The time evolution of $\Omega_{\rm GW}(\tau, k)$ is given by
\begin{equation}
	\Omega_{\rm GW}(\tau, k)=
	\frac{k^2\Delta_h^2(\tau,k)}{12a^2H^2}\,,
\end{equation}
where \mbox{$\Delta_h^2(\tau,k)=T_h(\tau,k){\cal P}_h(k)$}. The transfer function is given by 
\mbox{$T_h=\frac12(a_k/a)^2$}, where $a_k$ is the value of $a(\tau)$ at the moment of horizon re-entry of the scale with momentum $k$. Today we have \mbox{$T_h^0=\frac12(a_k/a_0)^2=\Omega_R
	\left(\frac{a_0H_0}{a_k H_k}\right)^2$}, where \mbox{$\Omega_R\simeq 10^{-4}$} is the density parameter of radiation at present and `0' denotes today.

Switching to frequency $f$, we employ the relation \mbox{$f=\frac{k}{2\pi}\frac{a_k}{a_0}$}. We end up with the expression \cite{OGWf}
\begin{equation}
	\Omega_{\rm GW}(f)\propto f^{-2(\frac{1-3w}{1+3w})}\,,
	\label{OGWf}
\end{equation} 
where $w$ is the barotropic parameter of the Universe at the time of horizon reentry. Thus, we see that, for modes which re-enter the horizon during the radiation era, because $w=\frac13$, we have
\mbox{$\Omega_{\rm GW}(f)=\,$constant}. That is, the spectrum is flat and unfortunately, its value is unobservable in the near future (see Fig.~\ref{fig}).

\section{Kination}

The inflationary paradigm suggests that the Universe inflates when dominated by the potential density of a scalar field, called the inflaton field. Non-oscillatory inflation considers a runaway inflaton scalar potential, with its minimum displaced at infinity \cite{NOinf}. 
Is such models, after the inflaton field $\phi$ rolls away from the inflationary plateau, which is a relatively flat part of its scalar potential $V(\phi)$, it becomes dominated by its kinetic energy density $\frac12\dot\phi^2$, because the potential reduces drastically and becomes negligible. The Universe is still dominated by the inflaton field, but the latter is kinetically dominated, with barotropic parameter \mbox{$w=\frac{\frac12\dot\phi^2-V}{\frac12\dot\phi^2+V}\approx 1$}. This phase is called {\em kination} \cite{kination}. 
Eq.~(\ref{OGWf}) suggests that, for the modes that re-enter the horizon during kination we have \mbox{$\Omega_{\rm GW}(f)\propto f$}.

Therefore, the GW spectrum features a peak. The corresponding frequencies, however, are unobservable at the moment because kination cannot be extended arbitrarily to later times, and therefore, lower frequencies. The reason is that, if kination is prolonged, the GW peak becomes too large and threatens to destabilise the delicate process of Big Bang Nucleosynthesis (BBN). The upper bound to $\Omega_{\rm GW}$ is obtained as follows.

At the time of BBN we require that \mbox{$\Omega^{\rm BBN}_{\rm GW}<10^{-2}$} so that BBN is not harmed. Using that the density of GWs redshifts as radiation we can estimate the corresponding bound at present. We find
\begin{equation}
	\Omega_{\rm GW}^0=\frac{\rho_{\rm GW}^0}{\rho_c^0}=\left.\frac{\rho_{\rm GW}}{\rho_r}\right|_0\Omega_R=\left.\frac{\rho_{\rm GW}}{\rho_r}\right|_{\rm BBN}\Omega_R=\Omega_{\rm GW}^{\rm BBN}\Omega_R<10^{-6}\,,
\end{equation}
where $\rho_r$ is the density of radiation and we used that \mbox{$\rho_{\rm GW}/\rho_r=\,$constant}. Thus, the sharp peak in $\Omega_{\rm GW}(f)$ of kination cannot be extended to observable frequencies (see Fig.~\ref{fig}).

\section{Stiff period}

If the peak in $\Omega_{\rm GW}(f)$ is not so sharp then it might be extended to observable frequencies without disturbing BBN. This may be possible if \mbox{$\frac13<w<1$}. Indeed, In Ref.~\cite{FT} it is shown that, when \mbox{$0.46\leq w\leq 0.56$} and \mbox{1~MeV$\,<T_{\rm reh}<150\,$MeV}, then the GW peak can be extended to frequencies low enough to be observable in the near future by Advanced LIGO and LISA, where $T_{\rm reh}$ is the reheating temperature, that is the temperature of the thermal bath at the onset of the usual radiation era of the hot Big Bang.

How can this possibility be realised? I have presented a concrete model to this end in Ref.~\cite{mine}. Consider two flat directions $\varphi$ and $\sigma$ in field space, which meet at an Enhanced Symmetry Point (ESP) such that 
they are characterised by the standard hybrid potential
\cite{hybrid}
\begin{equation}
	V(\varphi,\sigma)=\frac12 g^2\sigma^2\varphi^2+\frac14\lambda(\varphi^2-M^2)^2+V(\sigma)\,,
	\label{Vvarphi}
\end{equation}
where $g<1$ is a perturbative interaction coupling, $\lambda<1$ is a perturbative self-coupling and $V(\sigma)$ is some unknown potential for the inflaton field $\sigma$, which forces it to vary (roll) to smaller values. In the above $M$ is the vacuum expectation value (VEV) of the waterfall field $\varphi$. Below we consider that \mbox{$M\sim m_P$}, which is why $\varphi$ can be called a flat direction (only lifted by Planck-suppressed interactions).

The waterfall field is non-canonical. Indeed, the the Lagrangian density is 
\begin{equation}
	{\cal L}=-\frac12\partial_\mu\sigma \partial^\mu\sigma-
	\frac{\frac12\partial_\mu\varphi\partial^\mu\varphi}{(1-\varphi^2/M^2)^2}-
	V(\varphi,\sigma)\,,
\end{equation}
i.e. there are poles at the VEVs of $\varphi$ due to non-trivial geometry in field space. This is the standard setup in
$\alpha$-attractors, where the poles could be due to a non-trivial K\"{a}hler metric in supergravity \cite{alpha}. 
We can define a canonically normalised waterfall field $\phi$ using the transformation
\mbox{$\varphi=M\tanh(\phi/M)$}. In terms of $\phi$, the scalar potential in Eq.~(\ref{Vvarphi}) assumes the form
\begin{equation}
	V(\phi,\sigma)=\frac12g^2M^2\sigma^2
	\tanh^2(\phi/M)+\frac{\frac14\lambda M^4}{\cosh^4(\phi/M)}+V(\sigma)\,,
	\label{Vphisigma}
\end{equation}
where the minima along the waterfall direction have been displaced at infinity.

When the inflaton expectation value is large, 
the waterfall field is heavy and is pushed towards the origin. At the origin, $\varphi$ is canonical, so the hybrid mechanism operates normally. The waterfall transition occurs when the inflaton reaches the critical value $\sigma_c=(\sqrt\lambda/g)M$
\cite{hybrid}. Afterwards, the waterfall field finds itself on top of a potential hill and is released along its runaway direction towards large (absolute) values. 

%The runaway waterfall 
%direction is characterised by the potential
%\begin{equation}%
%	V(\phi)=\frac{\frac14\lambda M^4}{\cosh^4(\phi/M)}\,.
%	\label{Vphi}
%\end{equation}

Near the origin, when \mbox{$\phi\ll M$} (without loss of generality we assume $\phi>0$), the %above 
runaway waterfall 
potential is approximated as
\begin{equation}
	V\simeq\frac{\frac14\lambda M^4}{\left[1+\frac12(\phi/M)^2\right]^4}\simeq\frac14\lambda M^4\left[1-2\left(\frac{\phi}{M}\right)^2\right]\,.
\end{equation}	
Because $M\sim m_P$ (see below), the waterfall field undergoes a period of quadratic hilltop inflation, while $\phi$ dominates the Universe \cite{hilltop}. 

Eventually, $\phi\gg M$ and the waterfall
potential %in Eq.~(\ref{Vphi}) 
is approximated as
\begin{equation}
	V\simeq\frac{\frac14\lambda M^4}{\left[\frac12\exp(\phi/M)\right]^4}\simeq 4\lambda M^4 e^{-4\phi/M}\,.
\end{equation}
During this roll, there is an attractor solution (power-law inflation \cite{powerlaw}) in which the barotropic parameter of the rolling scalar field is
\begin{equation}
	w=-1+\frac{16}{3}\left(\frac{m_P}{M}\right)^2\,.
\end{equation}
Thus, the value \mbox{$M\approx 1.88\,m_P$} results in \mbox{$w\simeq\frac12$}, which means that there is a stiff period of the Universe history when the GW modes re-entering the horizon correspond to a peak with \mbox{$\Omega_{\rm GW}(f)\propto f^{2/5}$}
[cf. Eq.~(\ref{OGWf})], which is not as sharp as the one due to kination and can be extended to observable frequencies (see Fig.~\ref{fig}). 

A multiple of mechanisms can be responsible for reheating at the appropriate time. In Ref.~\cite{mine}, Ricci reheating is assumed as an example \cite{RicciReh}, where the Universe is reheated by the decay of a spectator field non-minimally coupled to gravity. It is shown that the appropriate $T_{\rm reh}$ is obtained with non-minimal coupling \mbox{$\xi\simeq 30$}.

\section{Hyperkination}

There is another example of creating enhanced primordial GWs by inflation, this time by truncating a peak in the GW spectrum generated by a stiff period. I have investigated this with collaborators in Ref.~\cite{ours}. It can be done as follows.

In Palatini modified gravity we consider
\begin{equation}  
	{\cal L}=\frac12 m_P^2R+\frac12\alpha R^2+\frac{1}{2}\xi\varphi^2 R -  \frac12\partial_\mu\varphi\,\partial^\mu\varphi - V(\varphi)\,,
\end{equation}
where $\alpha$ and $\xi$ are non-perturbative coefficients. Switching to the Einstein frame we obtain
\begin{equation}
{\cal L}=\frac12 m_P^2 R -\frac12\partial_\mu\phi\,\partial^\mu\phi + \frac14\alpha\frac{h^2+4\alpha V}{h^2m_P^4}(\partial_\mu\phi\,\partial^\mu\phi)^2-\frac{Vm_P^4}{h^2+4\alpha V}\,,
\end{equation}
where \mbox{$h(\varphi)= m_P^2+\xi\varphi^2$} and we employed the field redefinition 
\begin{equation}
	\frac{{\rm d}\phi}{{\rm d} \varphi}=\sqrt{\frac{hm_P^2}{h^2+4\alpha V}} \,. 
\end{equation}
In the above there is a strange quartic kinetic term. Such a term can be considered in general k-inflation models (no need for Palatini modified gravity) \cite{kinf}.

\begin{figure}
	\centering

	\includegraphics[scale=0.6]{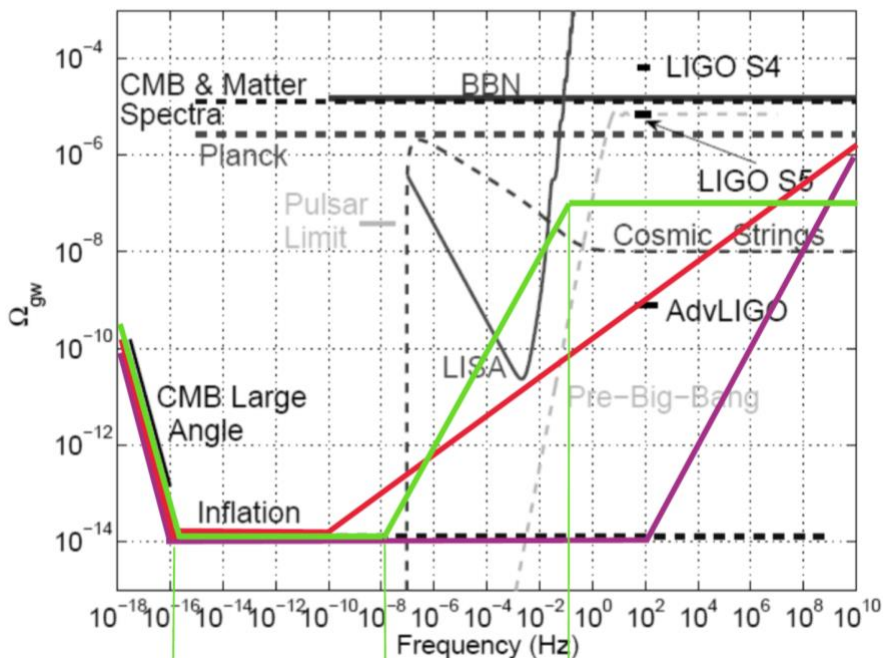}
	\vspace{-10cm}
	\caption{Plot of $\Omega_{\rm GW}(f)$ superimposed with the observational expectations of LISA and Advanced LIGO (taken by Ref.~\cite{GWobs}). The frequency range extends up to the scale of inflation, corresponding to the largest possible frequency, of GW which re-enter the horizon right at the end of inflation (assumed at the energy scale of grand unification). The uplift in the spectrum at low frequencies corresponds to the matter era of the hot Big Bang. The case when inflation is directly followed by the radiation era of the hot Big Bang (prompt reheating) is depicted by the low horizontal dashed thick line. As shown, the predicted spectrum is well beyond the observational capabilities of LISA and Advanced LIGO. The case when an early era of kination follows right after inflation corresponds to the purple line. As shown, there is a sharp peak in the spectrum (\mbox{$\Omega_{\rm GW}(f)\propto f$}), which however cannot be larger than the upper horizontal dashed line, which is the BBN constraint \mbox{$\Omega_{\rm GW}<10^{-6}$}. Thus, the kination peak cannot extend to low frequencies and is not near the expected observations of LISA and Advanced LIGO. The case of a stiff period with $w\approx\frac12$ following right after inflation is depicted with the red line. The peak of GW spectrum is milder (\mbox{$\Omega_{\rm GW}(f)\propto f^{2/5}$}), which means that it can be extended to lower frequencies without violating the BBN bound. As shown, the spectrum will be detectable by both LISA and Advanced LIGO. Finally, the case when a period of hyperkination (with a flat spectrum) and then regular kination follows the end of inflation is shown with a green line, which well overlaps with the expected observations of LISA and Advanced LIGO. In all cases, after reheating, the usual radiation era of the hot Big Bang begins and the GW spectra become flat as the frequency is lowered.}
	\label{fig}
\end{figure}

The EoM is
\begin{eqnarray}
\left[1+3\alpha\left(1+\frac{4\alpha V}{h^2}\right)\frac{\dot{\phi}^2}{m_P^4}\right]\ddot{\phi} + 3\left[1+\alpha\left(1+\frac{4\alpha V}{h^2}\right)\frac{\dot{\phi}^2}{m_P^4}\right] H \dot{\phi} & \nonumber\\
 +\;3\alpha^2\frac{\dot{\phi}^4}{m_P^4}\frac{\rm d}{{\rm d}\phi}\left(\frac{V}{h^2}\right) + \frac{\rm d}{{\rm d}\phi}\frac{Vm_P^4}{h^2+4\alpha V} & 
			= 0 \, .
\end{eqnarray}
Then, from the energy-momentum tensor we can obtain the energy density and pressure of the field, which read
\begin{eqnarray} 
		\rho_\phi &=& \frac{1}{2}\left[1+\frac{3}{2}\alpha\left(1+\frac{4\alpha V}{h^2}\right)\frac{\dot{\phi}^2}{m_P^4}\right]\dot{\phi}^2 + \frac{Vm_P^4}{h^2+4\alpha V} \, , \\
		p_\phi &=& \frac{1}{2}\left[1+\frac{1}{2}\alpha\left(1+\frac{4\alpha V}{h^2}\right)\frac{\dot{\phi}^2}{m_P^4}\right]\dot{\phi}^2 - \frac{Vm_P^4}{h^2+4\alpha V} \, .
\end{eqnarray}
After exiting the inflationary plateau, the inflaton field $\phi$ becomes dominated by its kinetic energy density, i.e. it becomes oblivious to the potential $V$. Then the above reduce to
\begin{equation}
	\left(1+3\alpha\frac{\dot{\phi}^2}{m_P^4}\right)\ddot{\phi} + 3\left(1+\alpha\frac{\dot{\phi}^2}{m_P^4}\right) H \dot{\phi} = 0 
\end{equation}
and 
\begin{equation}
	\rho_\phi = \frac{1}{2}\left(1+\frac{3}{2}\alpha\frac{\dot{\phi}^2}{m_P^4}\right)\dot{\phi}^2 \quad{\rm and}\quad
	p_\phi = \frac{1}{2}\left(1+\frac{1}{2}\alpha\frac{\dot{\phi}^2}{m_P^4}\right)\dot{\phi}^2 \, .
\end{equation}
Thus, when the quadratic kinetic term dominates one can effectively set $\alpha=0$ and we have regular kination with
\mbox{$w=1$}. However, when the quartic kinetic term dominates we can effectively consider only the $\alpha$-depended terms. Then, \mbox{$w=p_\phi/\rho_\phi=\frac13$}. We call this period {\em hyperkination}. Eq.~(\ref{OGWf}) suggests that the corresponding part of the GW spectrum is flat as in the radiation era. This means that the peak generated by kination has been truncated, which implies that the kinetic regime can last longer without disturbing BBN. As such, the GW signal can be amply boosted at observable frequencies as shown in Fig.~\ref{fig}.

\section{Conclusions}

Cosmic Inflation resolves the fine-tunings of the hot Big Bang and provides seeds for structure formation. Inflation is spectacularly verified by CMB observations. 
Another generic prediction of inflation is a superhorizon spectrum of primordial gravitational waves (GWs) generated through particle production.
The form of the resulting GW spectrum depends on the post-inflation history.
However, when GW modes re-enter the horizon during radiation domination they form a flat spectrum, too faint to be observable at present.

A stiff period in the Universe history enhances primordial GWs forming a peak in their spectrum. Non-oscillatory  inflation is followed by such a period, dominated by the inflaton’s kinetic energy density, called kination, but the frequencies of the peak are too high.
The GW peak can be extended to observable frequencies if the stiff period is milder than that of kination, with $w\approx 1/2$.
A model realisation of this possibility considers two flat directions which intersect at an ESP and give rise to the hybrid mechanism with Planckian waterfall VEV, which is also a kinetic pole of the waterfall field, as in $\alpha$-attractors.

Another possibility to obtain a boost in primordial GWs down to observable frequencies is by considering higher order kinetic terms, as with k-inflation. 
This is possible to realise in Palatini modified gravity. 
Considering $R+R^2$ gravity and a non-minimally coupled scalar field, results in additional quartic kinetic terms.
When the quartic kinetic terms dominate, this gives rise to to hyperkination.
Hyperkination is followed by regular kination, when the kinetic terms become canonical.
The resulting truncated GW peak can be extended to observable frequencies without disturbing BBN.

Forthcoming observations of Advanced LIGO, LISA, DesiGO and BBO may well detect the primordial GWs generated by inflation. Detection of primordial GWs will not only  confirm a prediction of inflation but offer tantalising evidence of the quantum nature of gravity, because the Bunch-Davis vacuum of virtual gravitons is assumed as an initial condition for the generation of GWs during inflation by particle production.

\paragraph{Acknowledgements}
This work was funded in part by STFC with the consolidated grant: ST/X000621/1.


\end{document}